\documentclass[twocolumn,printnumbers,amsmath,amssymb,aps,prl]{revtex4-1}
\usepackage{graphicx}
\usepackage{colordvi}
\usepackage{color}

\begin{document}
\title{Coupling between particle shape and long-range interaction in the high-density regime}

\author{Can-can Zhou$^{1,\dag}$, Hongchuan Shen$^{1,\dag}$, Hua Tong$^2$, Ning Xu$^3$ and Peng Tan$^{1, \ast}$}

\affiliation{$^1$State Key Laboratory of Surface Physics and Department of Physics, Fudan University, Shanghai 200433, People's Republic of China\\
$^2$ Department of Fundamental Engineering, Institute of Industrial Science, University of Tokyo, 4-6-1 Komaba, Meguro-ku, Tokyo 153-8505, Japan\\
$^3$CAS Key Laboratory of Soft Matter Chemistry, Hefei National Laboratory for Physical Sciences at the Microscale, and Department of Physics,
  University of Science and Technology of China, Hefei 230026, People's Republic of China}
\date{\today}


\begin{abstract}
{By using long-range interacting polygons, we experimentally probe the coupling between particle shape and long-range interaction. For two typical space-filling polygons, square and triangle, we find two types of coupling modes that predominantly control the structure formation. Specifically, the rotational ordering of squares brings a lattice deformation that produces a hexagonal-to-rhombic transition in the high density regime, whereas the alignment of triangles introduces a large geometric frustration that causes an order-to-disorder transition. Moreover, the two coupling modes lead to small and large “internal roughness” of the two systems, and thus predominantly control their structure relaxations. Our study thus provides a physical picture to the coupling between long-range interaction effect and short-range shape effect in the high density regime unexplored before.}
\end{abstract}

\maketitle
The shape anisotropy widely exists in molecules, colloidal particles, granules and even living cells. Its influence to structure ordering is of fundamental importance to solid designs in condensed matter physics, self-assemblies in material sciences and emergent behaviors in biological systems~\cite{wasio2014self,glotzer2007anisotropy,damasceno2012predictive,jones2010dna,jones2015programmable,li2005triangular,zhao2015shape,zhang2010collective,chen2017weak}. Coupling carefully-designed shapes with fine-tuned interactions brings many possibilities that produce various types of structures including plastic crystals~\cite{zhao2009frustrated,zhao2011entropic}, complex lattices~\cite{jones2010dna,jones2015programmable,nagaoka2018superstructures}, quasi-crystals~\cite{haji2009disordered,nagaoka2018single} and glass~\cite{zhao2015shape}. Besides static structures, the coupling between shape and interaction also strongly influences dynamic responses~\cite{han2006brownian,zheng2011glass,yunker2011rotational,li2019dynamics,shen2019universal}, and determines the system relaxation properties such as plasticity, frigidity and melting behaviors~\cite{heggen2010plastic,anderson2017shape,shen2019universal}.

Previous studies have highlighted the importance of shape effect when it couples with a short-range interaction~\cite{damasceno2012predictive,mao2013entropy,chen2011directed}. The geometric constraint from neighboring particles produces a short-range shape effect that predominantly controls the lattice structure. Taking the simple cases of square and triangle as examples, their unique rotational symmetries force them to tile up the space with two different lattice structures: square particles form a square lattice and triangle particles form a honeycomb lattice~\cite{anderson2017shape}.
The shape effect also drives the formation of tetratic phase and triatic phase upon lowering the density~\cite{wojciechowski2004tetratic,zhao2012local,anderson2017shape}.

Different from a short range interaction, coupling a long range interaction with the shape of particles presents a totally different scenario. Here, long range means that the cutoff length $R_{c}$ of the pair potential $u(r)$ is much larger than the particle size. A recent study has shown that square and triangle particles can form a universal state: the hexagonal plastic crystal with a lattice constant much larger than the particle size~\cite{shen2019universal}. In this case, the long-range interaction  contribution prevails the of the anisotropic effect on the total potential energy,  which drives the formation of 6-fold lattice structure, and the coupling produces a medium-range shape effect that determines the modes of structure relaxation.
However, there are still many open questions in the high density regime, where particles are close to each other and the shape effect becomes strong enough to compete with the long-range interaction effect.
For example, how does the system resolve the contradiction between the shape effect and the long-range interaction effect? What is the dynamic response?
These questions are helpful to understand pressure driven phase transitions in atomic and colloidal systems, where long-range interactions are present.
However, experimentally it is quite challenging to observe the structure formation and relaxation in the high density regime, owing to the difficulties of compressing the system to a large density while enabling in situ observations.

In this work, by using a model system composed by long-range interacting polygons (squares and triangles), we experimentally investigate the coupling between particle shape and long-range interaction in the high-density regime with a single-particle resolution. We find two types of coupling modes that predominantly control the structure formation and relaxation. Specifically, the rotational ordering of squares brings a lattice deformation that produces a hexagonal-to-rhombic transition in the high density regime, whereas the alignment of triangles introduces a large geometric frustration that causes an order-to-disorder transition. Moreover, the two coupling modes lead to small and large “internal roughness” of the two systems, and thus predominantly control their structure relaxations.

\begin{figure}[t]
    \includegraphics[width=8cm]{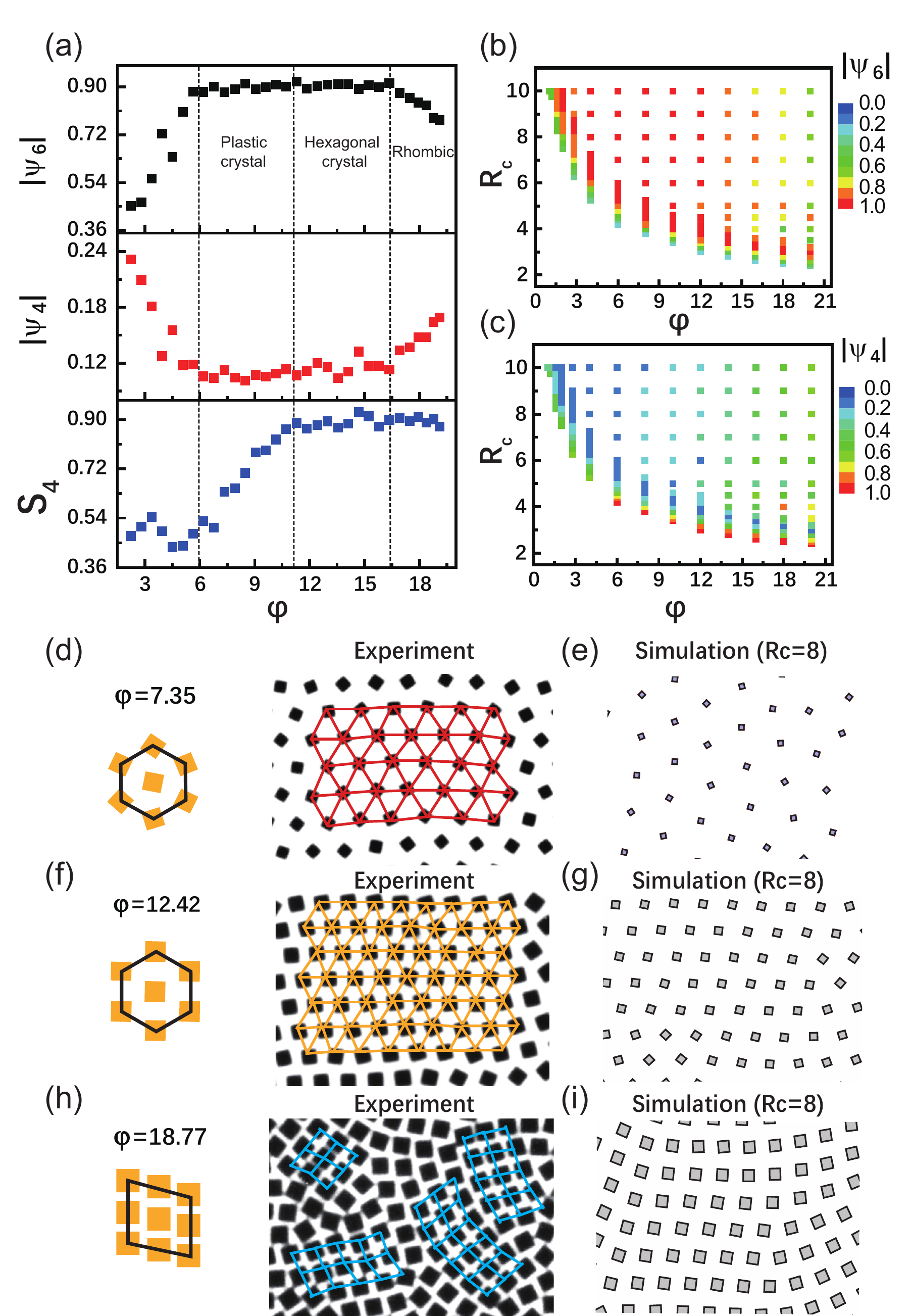}
    \caption{Hexagonal-to-rhombic structure transition of square-particle system in the high density regime. (a) Transitions illustrated by the bond orientational order parameter $\mid\psi_{6}\mid$ (top panel), $\mid\psi_{4}\mid$ (middle panel) and the rotational order parameter $S_{4}$ (bottom panel). Based on their evolutions with respect to the density $\varphi$, we can divide the bulk structures into hexagonal plastic crystal ($6<\varphi<11$), hexagonal crystal ($11<\varphi<17$) and rhombic structure ($\varphi>17$). On noting here we pick up top 50\% large value particles to calculate $S_{4}$ owing to the existence of point defects and grain boundaries. (b), (c) Illustration of the phase diagram from simulations.  A similar evolution of $\mid\psi_{6}\mid$ (upper panel) and $\mid\psi_{4}\mid$ (lower panel) as in the experiments can be observed at a cut off length $R_{c}\geq 5$. (d)-(i) Typical configuration of the three bulk structures from experiments and the agreements with simulations at $R_{c}=8$. Hexagonal lattice with a weak rotational order is shown in (d)($\varphi \sim 7.35$) and (e). Hexagonal lattice with long stripes of parallel aligned squares is shown in (f) ($\varphi \sim 12.42$) and (g). Rhombic structure is shown in (h) ($\varphi \sim 18.77$) and (i).}
    \label{fig1}
\end{figure}

We first explain our experimental systems. We use millimeter-sized magnets as particles, which have a polygon hard core (square and triangle, center-to-vertex distance $\sim 3$ $mm$) and a long-range magnetic interaction. The maximum of the magnetic field strength is about 0.2 $T$. We then confine the magnets in a two dimensional system. The interaction we measured in the confining plane is long-range repulsive (pair potential $u(r)\propto r^{-3.2}$ )~\cite{shen2019universal}. Significant potential anisotropy can be observed at small and medium $r$.
This system enables us to study the coupling between long-range interaction and the anisotropic shape effect up to the high density regime, where the lattice constant is close to the particle size. We keep the system size at a constant and gradually add more particles. The effective density $\varphi$ is defined in such a way that at about $\varphi$ = 1, the mutual repulsion just overcomes the friction and particles start to ``feel'' each other at a distance $r_{e}$. Here, $\phi=N\pi r_{e}^{2}/S$, with $N$ being the particle number and $S$ being the toal area. We equilibrate the system through small perturbations coming from a periodically moving plate, which has adjustable distance to the confining box and is randomly attached with magnets.

The system's structural order is quantified by such quantities: the n-fold bond orientational order parameter $|\psi_{n}|$ calculated from particle centers and the rotational order parameter $S_{n}$ calculated from the particle orientations. Here $|\psi_{n}|$ is the average of $\psi_{n,i}$, with $\psi_{n,i}=\frac{1}{N_{i}}\sum_{m=1}^{N_{i}}\exp^{in\theta_{mi}}$, where $N_{i}$ is the number of nearest neighbors around particle $i$, and $\theta_{mi}$ is the angle between  \textbf{$r_{m}$}-\textbf{$r_{i}$} and the $x$ axis. In comparison, a totally disordered state has $|\psi_{6}|=0$, a perfect hexagonal lattice has $|\psi_{6}|=1$, and a perfect square lattice has $|\psi_{4}|=1$.  The rotational order parameter $S_{n}$ is the average of $S_{n,i}$, which is defined as: $S_{n,i}=\frac{1}{N_{i}}\sum_{m=1}^{N_{i}}\cos(n\Theta_{m,i})$, with $\Theta_{m,i}$ being the angle difference between the orientation $\Theta_{m}$ of neighbor particle $m$ and the orientation $\Theta_{i}$ of centre particle $i$. A state with totally parallel aligned squares has $S_{4}=1$ and a state with random alignments has $S_{4}=0$. A state with totally parallel aligned triangles has $S_{3}=1$ and a honeycomb lattice that all the neighboring triangles are anti-parallel aligned has $S_{3}=-1$.

Based on evolutions of $|\psi_{6}|$, $|\psi_{4}|$ and $S_{4}$ with respect to the density $\varphi$ (Fig.~1a), we can clearly identify three types of bulk structures in square-particle systems: a hexagonal plastic crystal at $6<\varphi<11$, a hexagonal lattice structure at $11<\varphi<17$ and a rhombic structure at $\varphi>17$ (note that we pick up the first 50\% large value particles to calculate $S_{4}$ owing to the existence of point defects and grain boundaries). Typical configurations are separately illustrated in Figs.~1d, f, h.

The hexagonal plastic crystal is characterized by a high $|\psi_{6}|$ plateau and a continuous growth of $S_{4}$ (Fig.~1a).  It has a 6-fold lattice structure with many randomly aligned squares, as shown in Fig.~1d. It was explained by the long-range interaction in a previous study~\cite{shen2019universal}.

The hexagonal crystal is characterized by both a high $|\psi_{6}|$ and a high $S_{4}$ at $11<\varphi<17$. It implies that the anisotropic effect of the interaction becomes strong enough to produce a rotational ordering whereas the long-range effect can still retain a 6-fold positional ordering, as verified by the 6-fold lattice configuration composing of long stripes of parallel aligned squares shown in Fig.~1f.

The rhombic structure is characterized by a lowering of $|\psi_{6}|$, a growth of $|\psi_{4}|$ and a high value of $S_{4}$ at $\varphi>17$. It is caused by coupling of the rotational ordering with the 6-fold positional ordering at the large density, which deforms the hexagonal lattice to a rhombic one, as shown in Fig.~1h. Such a continuous deformation adjusts the confliction between the strong shape effect and the long-range interaction effect in the high density regime.

We confirm our experimental results through extensive simulations. The interaction that we used in Monte Carlo simulations has the form $u(r,\theta_{i},\theta_{j})=U_{0}/r^{-3.2}+ A(\theta_{i},\theta_{j})/r^{-4.7}$, with $A(\theta_{i},\theta_{j})=A_{0}[h(\frac{\theta_{i}}{\theta_{c}})+h(\frac{\theta_{j}}{\theta_{c}})+C$] being the anisotropic term, $h(x)= 1-3x^{2}+3x^{4}-x^{6}$, $\theta_{c}=\pi/4$ for squares and $\theta_{c}=\pi/3$ for triangles.  Here $U_{0}$, $A_{0}$ and $C$ are parameters that can adjust the interaction close to our experimental measurements~\cite{shen2019universal}. We use a cut-off distance $R_{c}$, in terms of the particle size (center-to-vertex distance), to control the interaction range. As shown by the phase diagrams in Figs.~1b-c, we universally observe a structure transition similar to our experimental observations when $R_{c}\geq 5$. The typical configurations in simulations agree well with experimental results, as shown in Figs.~1d-i. Therefore, we conclude that the rotational ordering of squares caused by the strong shape effect deforms the hexagonal lattice to a rhombic one at the high density regime. The generation mechanism of rhombic structures agrees with a previous study performed in hard-repulsive systems~\cite{zhao2011entropic}.

\begin{figure}[t]
    \includegraphics[width=8.5cm]{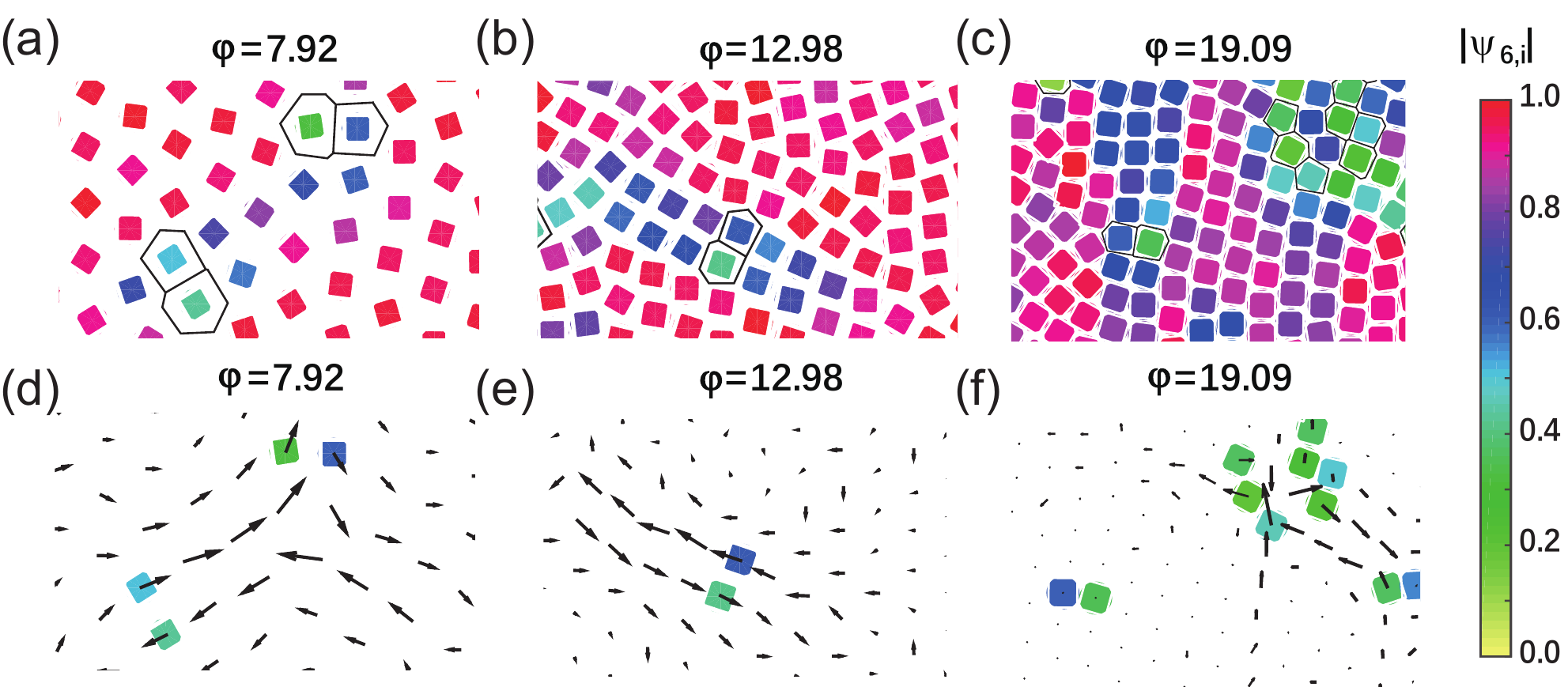}
    \caption{Structure relaxation dynamics in square-particle systems. (a)-(c) Three typical configurations formed through a structure relaxation event at three different densities: $\varphi\sim7.92$ in (a), $\varphi\sim12.98$ in (b) and $\varphi\sim19.09$ in (c). (d)-(f) Illustration of the displacement field during the structure relaxation corresponding to formation of the three configurations shown in (a)-(c). Gliding motion drives the structure relaxation at all densities.  During the structure relaxation, two nearby strips formed by parallel aligned squares moves in an opposite direction owing to the small ``internal roughness''.}
    \label{fig2}
\end{figure}

Interestingly, all the three structures shown in Fig.~1 have a similar structure relaxation mode. We observe dislocation gliding in hexagonal lattices at $6<\varphi<17$. Two nearby stripes of parallel aligned squares extensively move along an opposite direction, as shown by typical samples in Fig.~2a, d ($\varphi$=7.92) and Fig.~2b, e ($\varphi$=12.98).
For rhombic structure at $\varphi$=19.09, it is hard to distinguish the dislocation pairs. However, we still observe a similar structure relaxation mode that is featured by the large movement of stripes along the opposite direction, as shown in Fig.~2c, f. Such a common feature can be explained by the rotational ordering of squares, which produces stripes with small internal roughness and lubricate the large opposite motions. The explanations are also consistent with studies in hard cube systems, where the vacancies drive a sliding motion~\cite{smallenburg2012vacancy}.

Triangle particles in our system have a similar long-range interaction but a different shape effect compared with square particles. Consistently, we observe a hexagonal plastic lattice at $7<\varphi<9$, as characterized by a high $|\psi_{6}|$, a low absolute value of both $S_{3}$ and $S_{6}$ shown in Fig.~3a. At $\varphi>9$, the structure becomes more and more close to a glass state, as supported by a decrease of $|\psi_{6}|$ (upper of Fig.~3a) and a low saturated value of $S_{6}$ ($S_{6} \sim 0.12$) appearing at $\varphi>12$ (lower panel of Fig.~3a).
 The glass state can be explained by a confliction between the shape effect and the long-range interaction effect. Nearby triangles favor an anti-parallel alignment with two edges being close, which has a lower energy than the parallel alignment. It is supported by the negative value plateau of $S_{3}$
 ($S_{3}\sim -0.17$) appearing at $\varphi>12$, as shown in middle panel of Fig.~3a. However, the anti-parallel alignment can not be satisfied by all neighboring particles when the coordination number is large, as shown by three typical locally favored structures in Fig.~3c.
 The frustrated alignments generate many local polymorphs that may form a glass state at the high density regime~\cite{zhao2015shape,li2019dynamics}.

\begin{figure}[t]
  \begin{center}
    \includegraphics[width=8.5cm]{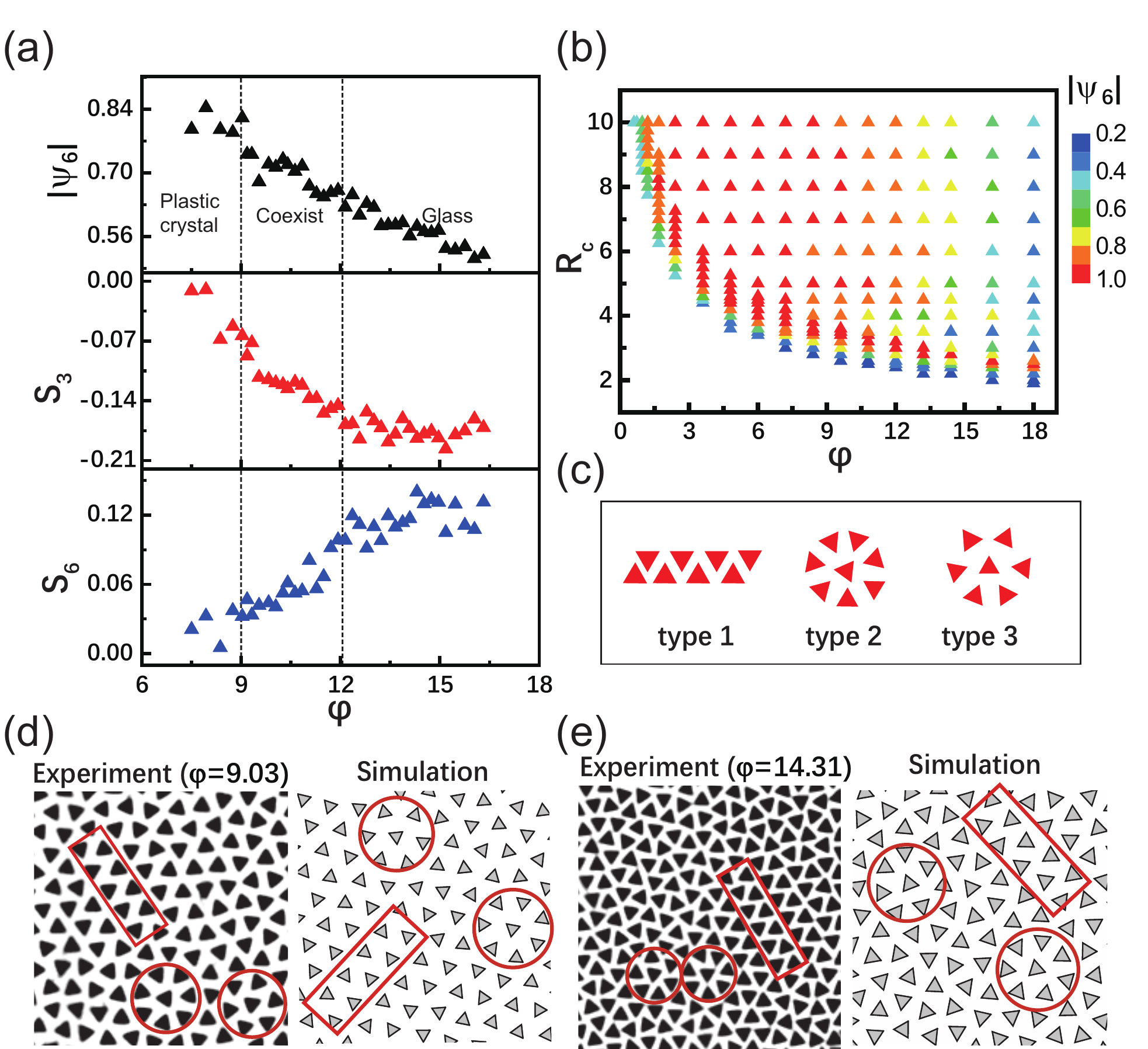}
    \caption{Order-to-disorder transition of triangle-particle systems in the high density regime. (a) Evolutions of $\mid\psi_{6}\mid$ (top panel), $S_{3}$ (middle panel) and $S_{6}$ (bottom panel). $\mid\psi_{6}\mid$ forms a high value plateau at $6<\varphi<9$ and starts to decrease as the further increase of $\varphi$. $S_{3}$ forms a plateau with negative value at $\varphi>12$. It implies a favor of the anti-parallel alignments of triangles. $S_{6}$ also forms a plateau with low positive value at $\varphi>12$. (b) Illustration of the phase diagram from simulations.  A similar evolution of $\mid\psi_{6}\mid$ as in the experiments can be observed at a cut off length $R_{c}\geq 5$. (c) illustration of typical local structures consisting of anti-parallel aligned particles. (d), (e) Typical configuration from experiments($\varphi \sim 9.3$ and $\varphi \sim 14.3$) and the agreements with simulations at $R_{c}=8$. Local structures shown in (c) are highlighted.}
  \label{fig3}
  \end{center}
\end{figure}

 Specifically, type 1 structure is composed of zigzag-connected anti-parallel aligned particle pairs. Type 2 structure contains one center particle and 7 neighbors that are roughly anti-parallel aligned. Type 3 structure consists of one center particle and 6 neighbors, with 4 of them are anti-parallel aligned. They produce a large geometric frustration that destroys the positional ordering at $\varphi>12$, as shown by two typical configurations in left panels of Fig.~3d ($\varphi=9.03$) and Fig.~3e ($\varphi=14.31$). Through extensive simulations, we find a similar order-to-disorder transition as in our experiments for $R_{c}\geq5$, as shown by the phase diagram in Fig.~3b, and two typical configurations in right panels of Figs.~3d-e ($R_{c}=8$). Our results suggest that an order-to-disorder transition can be produced by increasing the density of monodispersed and long-range interacting particles. Although the system may possibly form a complex lattice through repetitions of type 1 structure, the frustrated alignments and their numerous dynamic responses prevent such an ordering~\cite{li2019dynamics}. Our result enriches the families of the order-disorder transition, which are usually produced through changing the particle shape, size polydispersity and elastic heterogeneity~\cite{zhao2009frustrated,yunker2010observation,zheng2011glass,zhang2018compression}.

The frustrated alignment of triangle particles also intrinsically determines the structure relaxation mode in our system. We observe swirl-like large translational motions that strongly couple with large rotational motions, as shown by two typical samples at $\varphi=9.03$ (Figs.~a,c) and $\varphi=14.31$ (Figs.~4b,d). Interestingly, two nearby vortices ``rotate'' in the opposite direction (clock wise or anti-clockwise), which implies a large ``internal friction'' between them. Such a ``locked'' swirling motion can be explained by the large ``internal roughness'' in our system as a consequence of the large geometric frustration, which is a genuine feature of amorphous solids~\cite{cao2018structural,kou2018translational,zong2018vibrational,chen2010low,tan2012understanding,manning2011vibrational,widmer2008irreversible}.

\begin{figure}[t]
  \begin{center}
   \includegraphics[width=8.5cm]{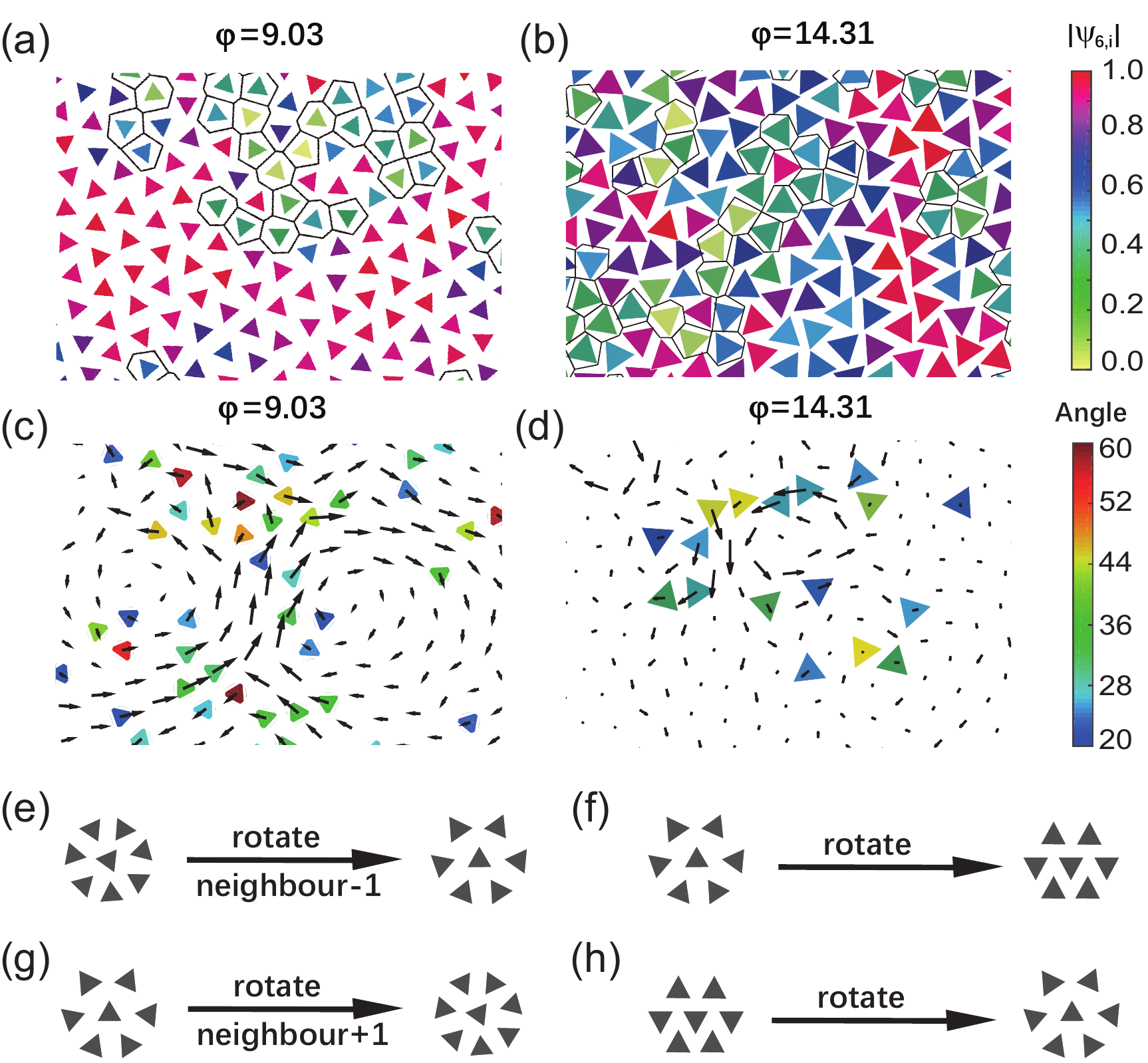}
    \caption{Structure relaxation dynamics in triangle-particle system. (a), (b) Typical configurations formed after a structure relaxation event at two different densities: $\varphi \sim 9.03$ in (a) and $\varphi\sim14.31$ in (b).  (c), (d) Illustration of the displacement field during the structure relaxation corresponding to the two configurations shown in (a) and (b). Swirl-like large translational motion couples to large rotational motion at all densities.  During the structure relaxation, two nearby swirls rotate in the opposite direction (clock wise or anti-clockwise) owing to the large internal roughness. (e)-(h) Illustration of local structure transformations. Type 3 serves as intermediate of structure transformation between type1 and type 2. The pathways form a closed loop.}
  \label{fig4}
  \end{center}
\end{figure}

Interestingly, the structure relaxations are mainly related to transformations among the three types of local structures shown in Fig.~3c. The transformation pathways are illustrated in Figs.~4e-h. Type 2 structure has a rather smooth border that can facilitate its rotational motion. With the rotation, one of the 7 neighboring particles is pushed out, producing type 3 structure, as illustrated in Fig.~4e. Type 3 structure has a unique feature: half of its border is smooth whereas the other half is rough. It may serve as intermediates of the structure transformations. With a small adjustment, type 3 can transform to a structure piece of type 1 that forming the type 1 array, as shown in Fig.~4f. We also find that the transformations are mutual(Figs.~4g,h), which form reversible pathways that can balance the population of these local structures.

Our observations may help to explain high-performance materials that have a high strength under a large deformation. High strength requires large internal roughness that reduces the mobility of soft defect spots whereas large deformation usually creates many defects that produce destructive structure relaxations. The balanced structure transformation pathways here promote none destructive structure relaxations and hence could solve the conflict.

In summary, we observe two coupling modes between particle shape and long-range interaction in the high density regime.
Parallel alignment of squares forms a rhombic structure, whereas frustrated alignment of triangles produces a glass state.
To emphasize, the coupling not only controls the structure formation but also governs the ``internal roughness'' that determines the mode of structure relaxation.

Many previous studies have revealed the couplings between the short-range shape effect and the short-range interactions.
However, both the range of shape effect and the nature of interaction can be adjusted, thereby creating many possibilities of coupling them together~\cite{ou2019kinetic}.
Our study here reveals the importance of the range of the interaction. It was shown that soft potential systems, for example harmonic or Hertzian, have various solid structures adjusted by the density~\cite{zu2017forming}. It is interesting to extend our studies to these systems, where the coupling should produce new types of structures.

{\bf Acknowledgments.}

This work is supported by National Natural Science Foundation of China under Grants Nos. 11774059 and 11734014.

$^\dag$C. Z. and H. S. contributed equally to this work. \\
$^\ast$tanpeng@fudan.edu.cn

%

\end{document}